\documentclass[fleqn,usenatbib]{mnras}

\usepackage{newtxtext,newtxmath}

\usepackage[T1]{fontenc}

\DeclareRobustCommand{\VAN}[3]{#2}
\let\VANthebibliography\thebibliography
\def\thebibliography{\DeclareRobustCommand{\VAN}[3]{##3}\VANthebibliography}

\usepackage{graphicx}	
\usepackage{amsmath}	

\title[A closer look at individual collisions of dust aggregates]{A closer look at individual collisions of dust aggregates:\\Material mixing and exchange on microscopic scales}

\author[S. Krijt et al.]{
Sebastiaan Krijt,$^{1}$\thanks{E-mail: s.krijt@exeter.ac.uk}
Sota Arakawa,$^{2}$
Mark Oosterloo$^{3}$
Hidekazu Tanaka$^{4}$
\\
$^{1}$Department of Physics and Astronomy, University of Exeter, Exeter, EX4 4QL, UK\\
$^{2}$Yokohama Institute for Earth Sciences, Japan Agency for Marine-Earth Science and Technology, 3173-25 Showa-machi, Kanazawa-ku, Yokohama, 236-0001, Japan\\
$^{3}$Kapteyn Astronomical Institute, University of Groningen, Landleven 12, 9747 AD Groningen, Netherlands\\
$^{4}$Astronomical Institute, Graduate School of Science, Tohoku University, 6-3 Aramaki, Aoba-ku, Sendai, 980-8578, Japan
}

\date{Accepted 2024 September 25. Received 2024 September 18; in original form 2024 July 24}

\pubyear{\the\year{}}

\begin{document}
\label{firstpage}
\pagerange{\pageref{firstpage}--\pageref{lastpage}}
\maketitle

\begin{abstract}
Collisions between aggregates with different histories and compositions are expected to be commonplace in dynamically active protoplanetary discs. Nonetheless, relatively little is known about how collisions themselves may contribute to the resulting mixing of material. Here we use state-of-the-art granular dynamics simulations to investigate mixing between target/projectile material in a variety of \emph{individual} aggregate-aggregate collisions, and use the results to discuss the efficiency of collisional mixing in protoplanetary environments. We consider sticking collisions (up to 10-20 m/s for our set-up) and disruptive collisions (40 m/s) of BPCA and BCCA clusters, and quantify mixing in the resulting fragments on both individual fragment and sub-aggregate levels. We find that the mass fraction of material that can be considered to be `well-mixed' (i.e., locally made up of a mix of target and projectile material) to be limited, typically between 3-6\% for compact BPCA precursors, and increasing to 20-30\% for more porous BCCA clusters. The larger fragments produced in disruptive collisions are equally heterogeneous, suggesting aggregate-aggregate collisions are a relatively inefficient way of mixing material with different origins on small scales.
\end{abstract}




\section{Introduction} \label{sec:intro}

Growth via agglomeration in gentle pair-wise collisions constitutes the first step in the planet formation process \citep[e.g.,][]{ Drazkowska2023, Krijt2023}. This process connects the (sub-)micron-sized grains and condensates that we see in e.g., scattered light images to the larger macroscopic `pebbles' seen in ALMA continuum images that are also the precursors of planetesimals in the streaming instability plus gravitational collapse scenario \citep{Johansen2014}.

Solids in discs also undergo complex dynamical evolution through a combination of coherent/orderly effects (e.g., settling and radial drift) and more random ones (turbulent/diffusion) \citep[e.g.,][and many others]{Ciesla2012, Misener2019, Oosterloo2023}. Additionally, local variations in composition (e.g., ice content) can result from variations in sublimation and condensation on grains of different sizes, particularly close to snowlines or following an accretion outburst of the central star \citep[e.g.,][]{Ros2013, okamoto2022,  HougeKrijt2023, Oosterloo2024}. As a result, aggregates with distinct compositions and/or dynamical histories are expected to be present in the same disc region to undergo collisions. Nonetheless, most astrophysical simulations following dust composition and coagulation generally assume perfect mixing of composition during collisions \citep[e.g.,][]{Krijt2016, Oosterloo2023, homma2024}.

While considerable efforts have been made to understand the possible outcomes of collisions in terms of growth, fragmentation, bouncing etc., as functions of particle size, velocity, as well as composition (e.g., silicates vs. ice) \citep[see][and other]{DominikTielens1997, Wada2009, Wada2013, Ringl2012, Hasegawa2021, Hasegawa2023, Arakawa2022, Arakawa2023}, less work has been done on understanding how the composition of the collision products varies and to what extent it reflects the original bodies. A visual inspection of snapshots of simulations of individual collisions, however, such as Fig.~1 of \citet{Hasegawa2021}, indicates that even at high velocities (44 m/s in this case), the mixing of materials from the two colliders is quite limited.

In this work, we set out to utilize well-tested granular dynamics simulations (Sect.~\ref{sec:numerical_methods}) to investigate material mixing in individual collisions in intermediate and high-velocity collisions as a way of studying collisional mixing in protoplanetary discs. In Sects.~\ref{sec:BPCA} and \ref{sec:BCCA} we focus on individual collisions and seek to quantify: (i) the compositional variation between fragments (mostly relevant for high-velocity disruptive events) and (ii) the degree of internal mixing on the microscopic/sub-aggregate scale for a variety of collisions. We then use these results to estimate the efficiency of mixing in protoplanetary discs in Sect.~\ref{sec:discussion}.

\begin{figure*}
\includegraphics[width=1.\textwidth]{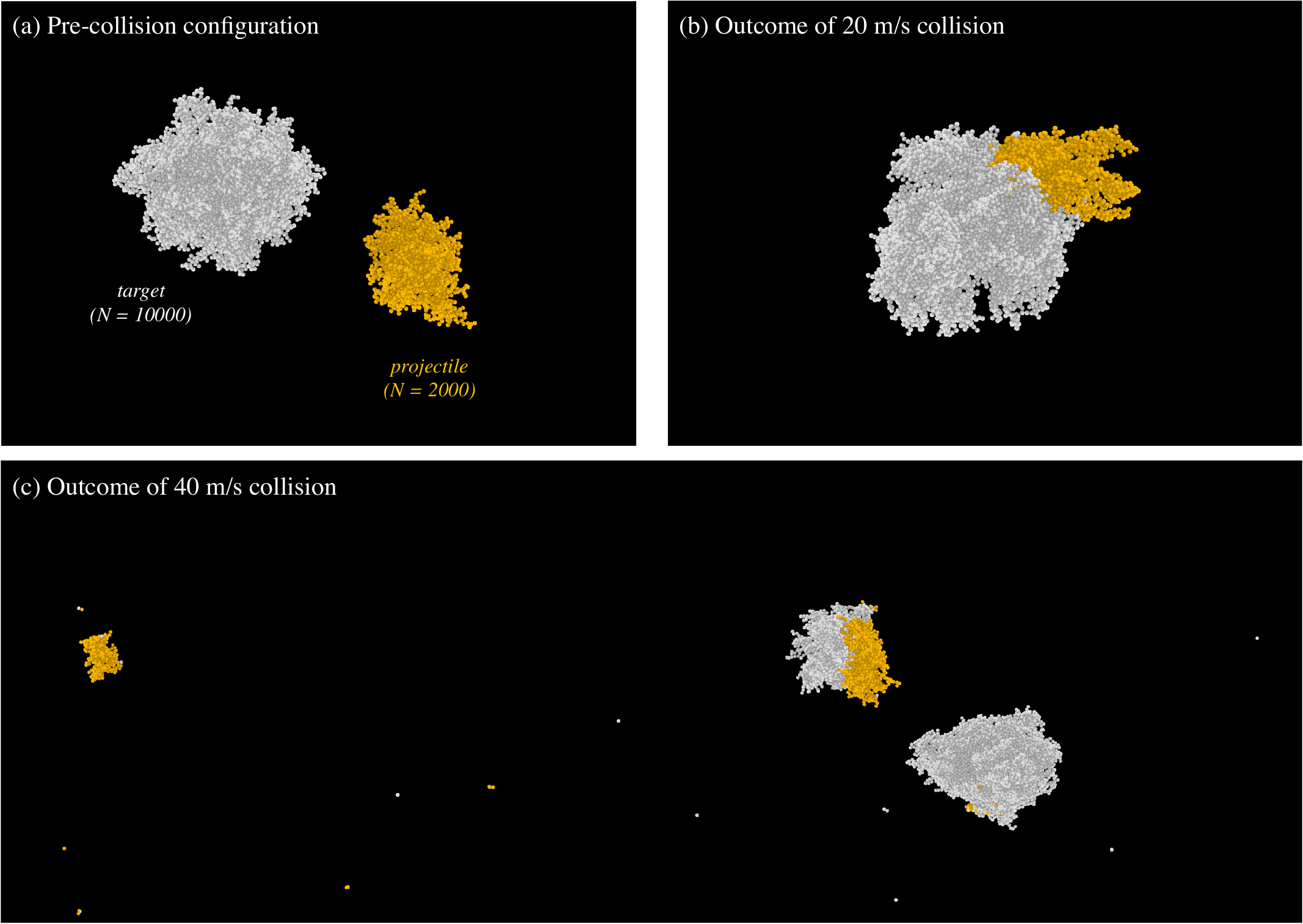}
\caption{(a) Pre-collision configuration showing the BPCA target (white) and projectile (gold) aggregates. (b) Outcome of a collision at $v_\mathrm{rel} = 20~\mathrm{m/s}$ that resulted in perfect sticking. (c) Outcome of a collision at $v_\mathrm{rel} = 20~\mathrm{m/s}$ that resulted in the creation of three large fragments and distribution of smaller ejecta. 
\label{fig:collage}}
\end{figure*}

\section{Numerical methods} \label{sec:numerical_methods}

We employ granular dynamics simulations similar to the ones described in \citet{Arakawa2022}, the approach being centered on a treatment of various interparticle forces, the description of which is based on the seminal work of \citet{DominikTielens1997, wada2007} and described in more detail below.

\subsection{Treatment of interparticle forces}

We perform three-dimensional simulations of collisions between two aggregates made of monodisperse ice particles (`monomers'). We focus on icy particles as icy dust represents the majority of the solid reservoir in protoplanetary discs and because the regions around icelines have be shown to be particularly sensitive to `non-local' processes such as dust transport via diffusion \citep{Oosterloo2024}. The radius of each monomer is set to be $r_\bullet = 100~\mathrm{nm}$. The material parameters of ice are assumed to be identical to those used in previous studies \citep[e.g.,][]{wada2007, Arakawa2022}; the material density is $\rho_\bullet = 1~\mathrm{g~cm^{-3}}$, Young's modulus is $7~\mathrm{GPa}$, Poisson's ratio is $0.25$, and the surface energy is $0.1~\mathrm{J~m^{-2}}$.

Monomer--monomer interactions for both normal and tangential motions are considered. We assume that ice monomers can be regarded as elastic spheres with cohesive force, and the inter-particle forces in the normal direction is modeled by \citet{jkr1971}. The tangential motion of two particles in contact is divided into three types: rolling, sliding, and twisting \citep[see][]{DominikTielens1997,wada2007}. The resistances against tangential displacements are modeled by linear elastic springs with inelastic thresholds \citep[e.g.,][Fig.~3]{wada2007}. The spring constants and critical displacements for tangential motions are set to be equal to those of previous studies \citep[][]{Hasegawa2021,Arakawa2022}.

\subsection{Preparation of BPCA and BCCA aggregates}

We consider two types of aggregation processes as preparing procedures of initial aggregates: ballistic particle-cluster aggregation (BPCA) and ballistic cluster-cluster aggregation (BCCA). BPCA aggregates are made by sequential hit-and-stick collisions between a cluster (aggregate) and single particle, and BCCA aggregates are made by sequential hit-and-stick collisions between two identical clusters with randomly chosen offsets and orientations \citep[][Fig.~3]{okuzumi2009}. These two types of aggregates are commonly used in numerical simulations of dust coagulation \citep[e.g.,][]{Wada2009, seizinger2013} and represent analogues of protoplanetary dust growing via gentle Brownian-motion-driven collisions with similar-size collision partners (for BCCA) and dust growing through sweep-up of smaller grains due to differential vertical settling or radial drift (for BPCA) \citep[e.g.,][]{zsom2011, Okuzumi2012}.

\subsection{Collisions considered}\label{sec:collisions_considered}

Next, we simulate collisions between a target (t) and projectile (p) with masses $M_\mathrm{t} = N_\mathrm{t} m_\bullet$ and $M_\mathrm{p} = N_\mathrm{p} m_\bullet$ where $N_i$ indicates the number of monomers in a given aggregate and $m_\bullet$ is the monomer mass (and $N_\mathrm{p} \leq N_\mathrm{t}$ by definition). We consider mass ratios of $N_\mathrm{p}/ N_\mathrm{t}$ of 1/2 (for BCCA) and 1/5 (for BPCA), similar to ratios that dominate growth of aggregates in protoplanetary disks where, except for the largest aggregates, collision rates between exactly same-sized aggregates are suppressed, see \citet[][Fig.~9]{Okuzumi2012} and \citet[][Fig.~10]{Krijt2015}. The pre-collision configuration for BPCA collisions is shown in Fig.\ref{fig:collage}(a), and the one for BCCA collisions in Fig.~\ref{fig:collage_bcca}(a). The characteristic radii, defined as $r_c = \sqrt{5/3} r_\mathrm{gyr}$ with $r_\mathrm{gyr}$ the gyration radius \citep{okuzumi2009}, of the target and projectile equal $42 r_\bullet$ and $25 r_\bullet$ for the BPCA and $130 r_\bullet$ and $105 r_\bullet$ for the BCCA aggregates, respectively, giving the bodies physical diameters between $5$ and $20~\mathrm{\mu m}$. 

In what follows we investigate collision velocities between 10 and 40 m/s. These velocities are close to the highest collision velocities typically found in turbulent protoplanetary discs, although particles below ${\sim}\mathrm{mm}$ sizes typically collide at lower velocities \citep[][Fig.~5]{birnstiel2023}. We choose these high velocities purposefully to cover both sticking and fragmentation regimes, and to avoid the `hit-and-stick' regime at low velocities where essentially no mixing takes place \citep[][]{blumwurm2008}. 

Finally, we consider two different impact parameters in this study. For the BPCA clusters we consider off-set collisions as these are the most common in protoplanetary discs. Specifically, we consider an impact parameter $b = 37 r_\bullet \approx 0.55 b_\mathrm{max}$, where $b_\mathrm{max}$ is the maximum impact parameter that can be achieved given the sizes of the target and impactor \citep[see][]{Wada2013}. In contrast, we focus on head-on ($b=0$) impacts for the BCCA clusters as a way of exploring the optimum scenario for mixing.

\subsection{Quantification of mixing}\label{sec:quantifying_mixing}

We consider mixing on the scale of whole aggregate fragments and on the scales of the interiors of individual fragments. First, to discuss mixing on the whole-fragment scale, i.e., considering aggregates and fragments as a whole, we define the \emph{projectile fraction} $\chi$ of a fragment, which is the mass fraction taken up by material originally associated with the projectile. Thus defined, the original target has $\chi=0$, the original projectile $\chi=1$, and the result of a perfect merger would have 
\begin{equation}
\chi = \chi^\mathrm{perf} = \frac{ M_\mathrm{p} }{(M_\mathrm{p}+M_\mathrm{t})} = \frac{N_\mathrm{p}}{(N_\mathrm{p}+N_\mathrm{t})}.
\end{equation}

Additionally, it is interesting to consider to what extent the interior of the collision product becomes mixed/homogenized, in particular in low-velocity sticking collisions. For this purpose, we define a \emph{local projectile fraction} $\chi^\ell(\vec{r})$ which corresponds to the projectile fraction inside a spherical volume with radius $\ell$ centered on position $\vec{r}$. Inside a \emph{perfectly homogeneously mixed} aggregate $\chi^\ell(\vec{r}) \approx \chi$ everywhere, while fragments that have not mixed internally will contain large regions where $\chi^\ell$ is either close to 0 or 1 as the local material is dominated by monomers originating from either the target or projectile, respectively. In the remainder of this work we will use $\ell = 5 r_\bullet = 0.5\mathrm{~\mu m}$, for which we found the local volume to be large enough to include a substantial amount of neighbours but still small compared to aggregate sizes to resolve sub-aggregate-level variations in composition. We discuss the impact this choice for $\ell$ has on our results in Appendix~\ref{sec:appendix_ell}.

\section{Collisions of BPCA Clusters}\label{sec:BPCA}

\subsection{Sticking collision at 20 m/s}\label{sec:20ms}
Figure~\ref{fig:collage}(b) depicts the outcome of a sticking collision at $v_\mathrm{rel} = 20~\mathrm{m/s}$. This collision resulted in perfect sticking without the loss of a single monomer, and thus the mass of the final product equals $M_\mathrm{f} = M_\mathrm{p} + M_\mathrm{t}$ and macroscopically the outcome is a perfect merger in terms of composition: $\chi= \chi^\mathrm{perf} = 2000/12000=1/6$.

\begin{figure}
\includegraphics[width=1.\columnwidth]{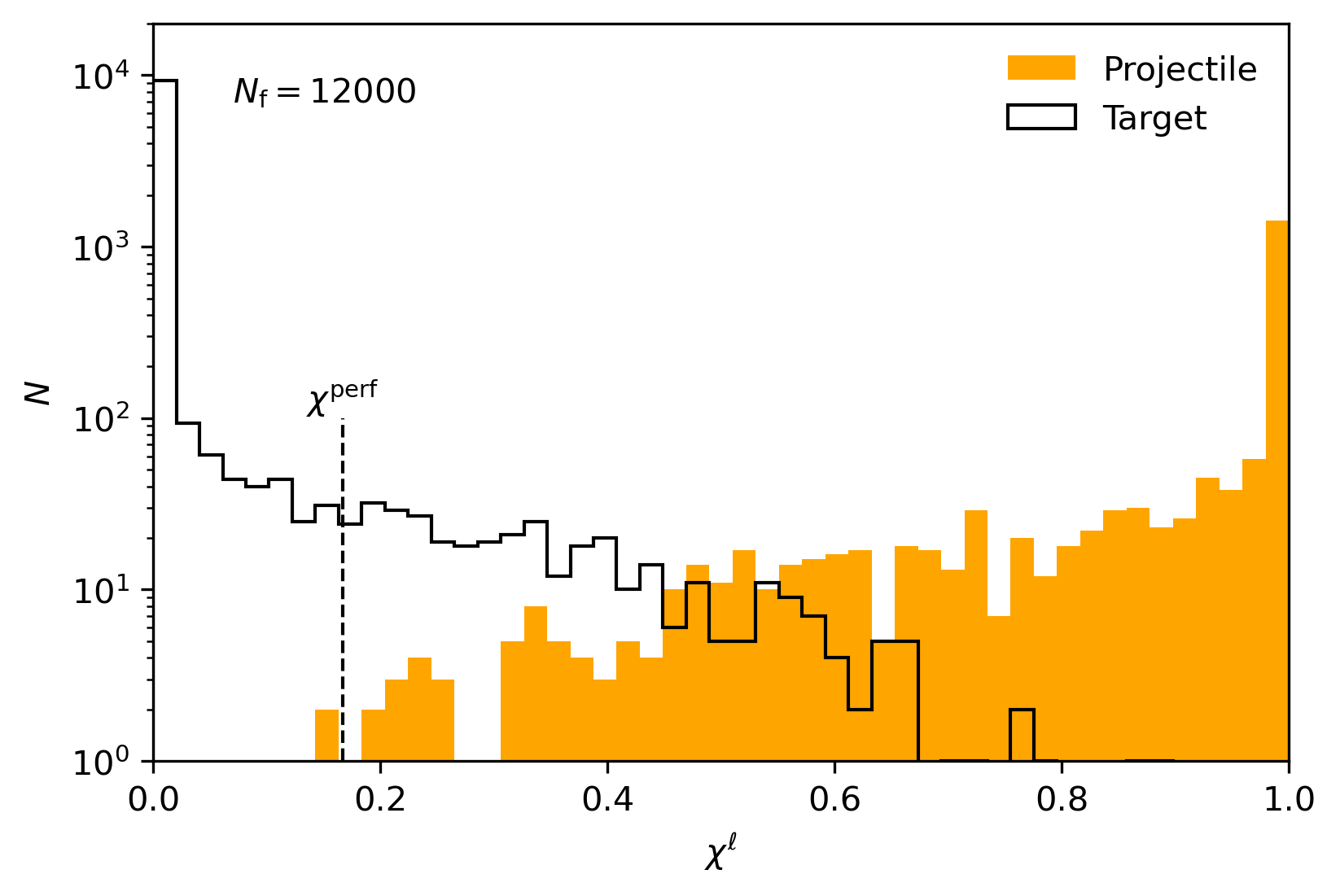}
\caption{Variation in the local projectile fraction of the aggregate depicted in Fig.~\ref{fig:collage}(b). The dashed line corresponds to $\chi^\ell = \chi^\mathrm{perf} = 1/6$ (see Sect.~\ref{sec:quantifying_mixing}). The two peaks at $\chi^\ell = 0$ (regions of pure target material) and $\chi^\ell = 1$ (pure projectile material) contain, respectively, 93\% of the original target and 70\% of the original projectile.
\label{fig:20ms_chi_ell}}
\end{figure}

However, examining Fig.~\ref{fig:collage}(b) it is clear that the projectile and target have not become mixed on the sub-aggregate level. To quantify this we show in Fig.~\ref{fig:20ms_chi_ell} histograms of the occurrence of different values of $\chi^\ell(\vec{r})$ assessed at each monomer location, distinguishing between monomers initially housed in the target and projectile. The distribution in Fig.~\ref{fig:20ms_chi_ell} shows two clear peaks at $\chi^\ell = 0$ (containing 93\% of the original target mass) and $1$ (containing 70\% of the original projectile mass), representing those regions dominated by material sourced entirely from one progenitor. The remaining $1287$ monomers (${\approx}10\%$ of the total mass involved in the collision) occupy the central part of the plot with intermediate values of $\chi^\ell$.

\begin{figure}
\includegraphics[width=1.\columnwidth]{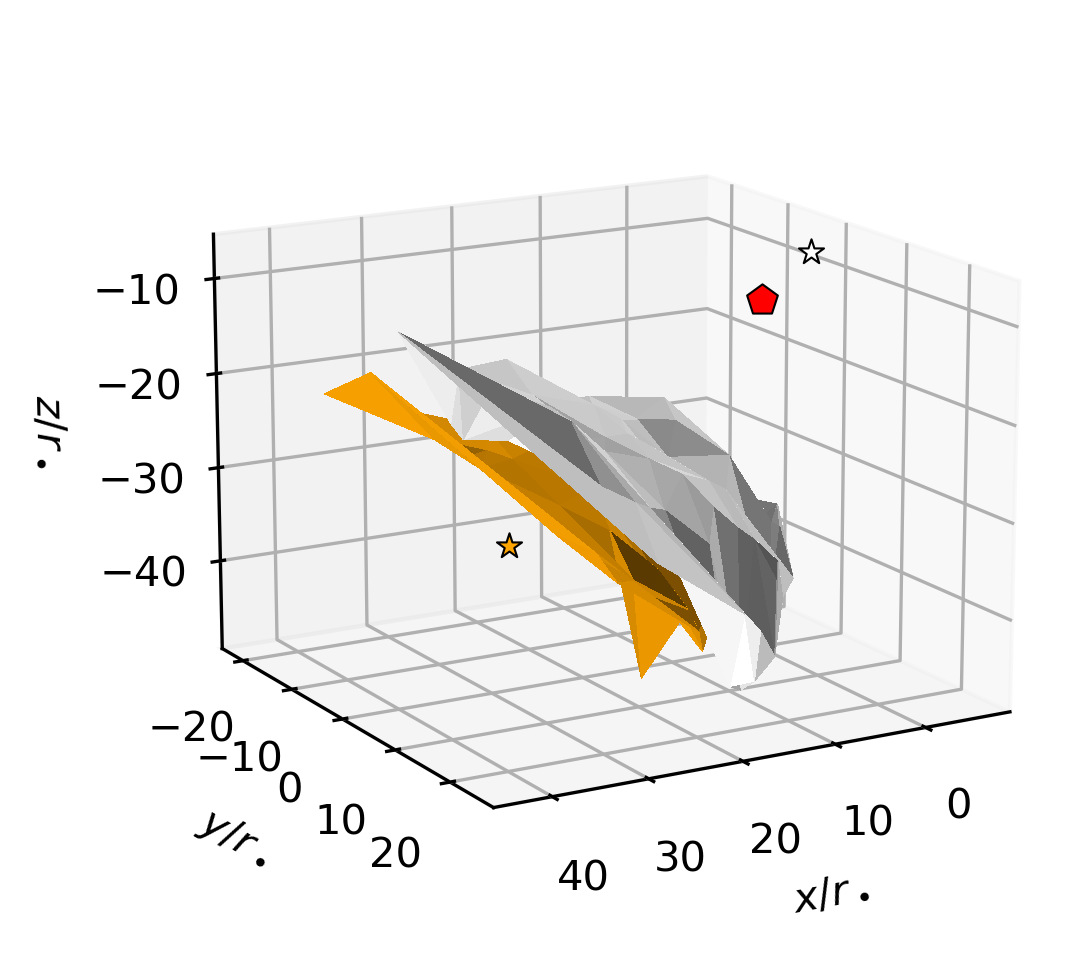}
\caption{Boundaries of surfaces corresponding to $\chi^{\ell}=0.2$ (white) and 0.8 (orange) shown relative to the center of mass of the aggregate (red pentagon) of the aggregate depicted in Fig.~\ref{fig:collage}(b). Orange and white stars denote the centers of mass of those portions originally belonging to the target and projectile, respectively.
\label{fig:20ms_interface}}
\end{figure}

To further illustrate the lack of internal mixing, we show in Fig.~\ref{fig:20ms_interface} the internal surfaces corresponding to $\chi^{\ell}=0.2$ (white) and 0.8 (orange). Only the volume in between these sheets can be considered well-mixed. The offset between the centers of mass (COM) of the portions originating in the target and projectile further highlight the lack of mixing.

\begin{figure}
\includegraphics[width=1.\columnwidth]{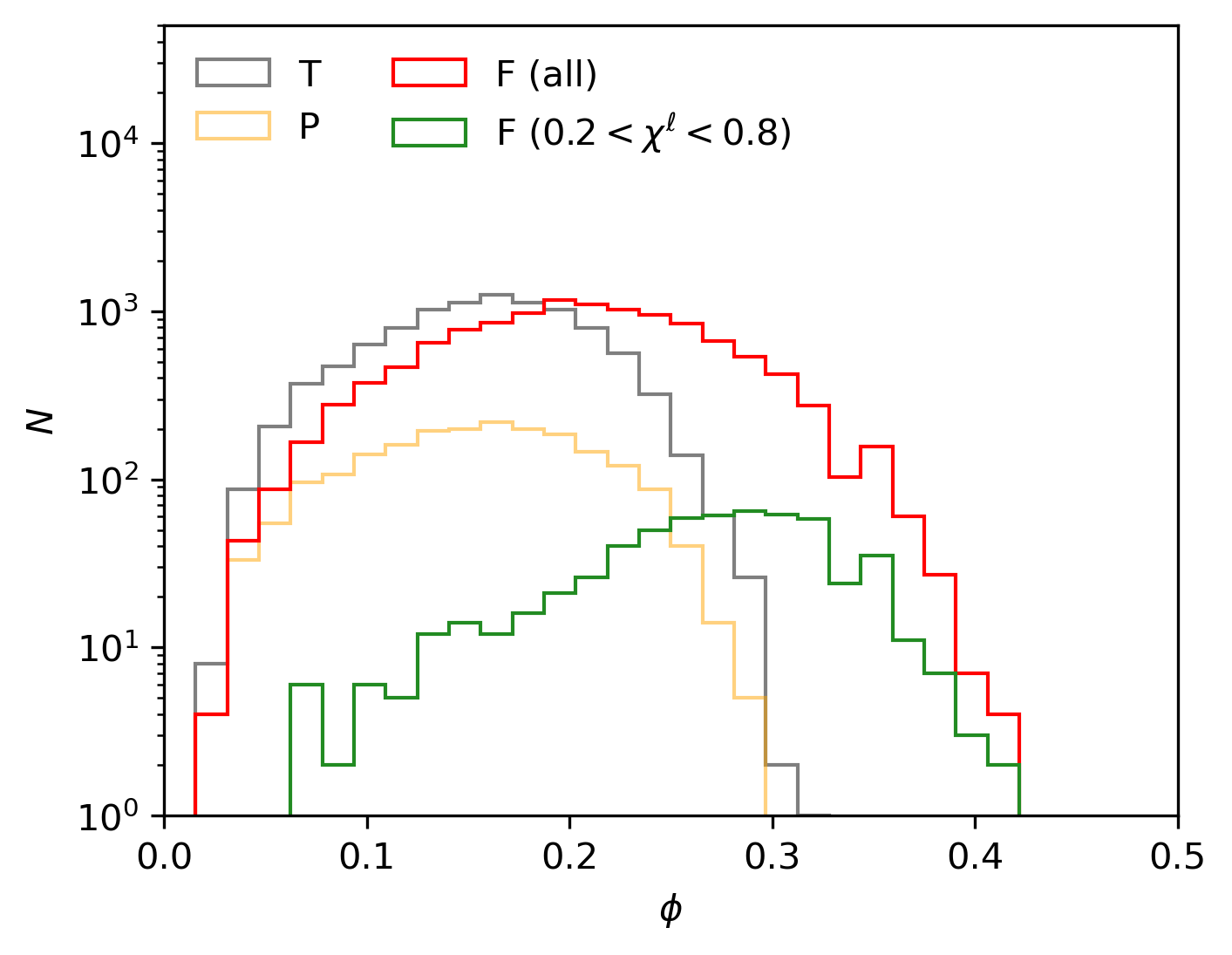}
\caption{Mass-weighted distributions of internal filling factor $\phi$ in the pre-collision BPCA target (T), BPCA projectile (P), post-collision fragment formed in a 20 m/s sticking collision (F), and that portion of the fragment containing a mixture ($0.2 < \chi^\ell < 0.8$)  of the original target and projectile (green).
\label{fig:20ms_ff}}
\end{figure}

Finally, Fig.~\ref{fig:20ms_ff} shows the distribution of internal filling factors for the target and projectile prior to the collision (in gray and orange), and the formed collision fragment in red. These distributions are produced using the following procedure: For every monomer that is part of the aggregate in question, we again consider a spherical volume with radius $\ell$ centred on that monomer's position $\vec{r}$. We count the number of monomers $N^\ell$ contained in that volume\footnote{For simplicity we do not consider partial monomers and simply count the number of monomers with position vectors $|\vec{r}_i - \vec{r}| \leq \ell$.}, and then compute the local filling factor as the ratio of the volume taken up by the $N^\ell$ monomers compared to the total spherical volume
\begin{equation}\label{eq:filling_factor}
\phi(\vec{r}) = N^\ell \left( \dfrac{ r_\bullet}{\ell} \right)^3.
\end{equation}
Defined in this way and evaluated once at every monomer position, the curves in Fig.~\ref{fig:20ms_ff} effectively represent a mass-weighted local filling factor distribution and cannot be directly compared to the aggregate-averaged porosity as used in e.g., \citet{okuzumi2009, Okuzumi2012}, which is defined using an aggregate's gyration radius.

For this particular 20 m/s sticking collision, Fig.~\ref{fig:20ms_ff} shows the impact led to an increase in internal density and the creation of a considerable region with filling factors $\phi \sim 0.3{-}0.4$ which were absent in the pre-cursors. The region of the new aggregate that consists of a mixture of target and projectile ($0.2 < \chi^\ell < 0.8$) predominantly consists of this more compressed material, its filling factor distribution peaking at 0.3.

\begin{figure}
\includegraphics[width=1.\columnwidth]{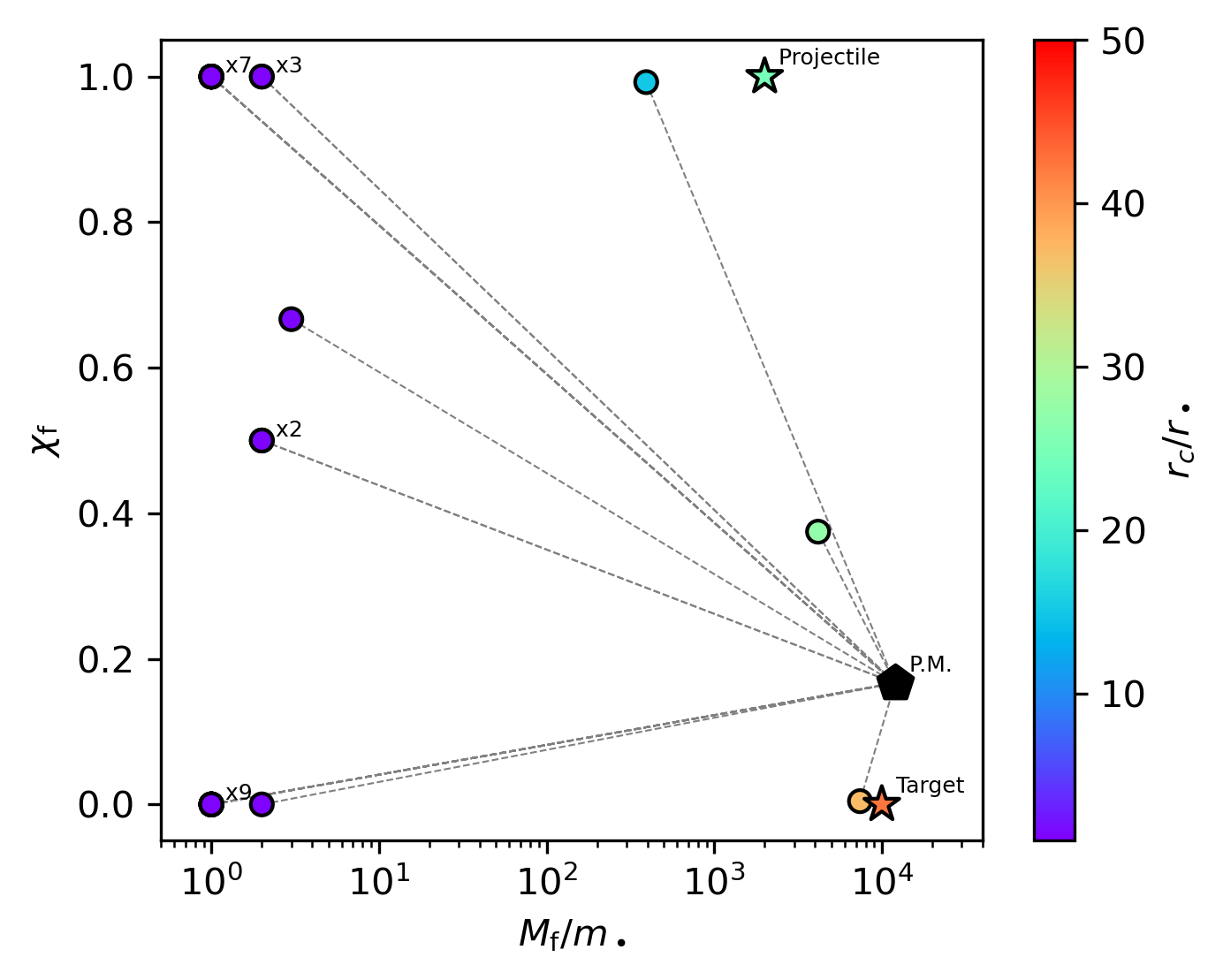}
\caption{Circles: Fragment masses and (macroscopic) projectile fractions ($\chi_\mathrm{f}$, see Sect.~\ref{sec:quantifying_mixing}) of the fragments produced in the 40 m/s BPCA collision depicted in Fig.~\ref{fig:collage}(c). Numbers denote the number of duplicates produced (for the small fragments), the star symbols correspond to the original target and projectile, and the colourbar indicates each fragment's characteristic size $r_c$. The black pentagon indicates the mass and $\chi_\mathrm{f}$ that would result from the perfect merger of all fragments (symbolized by the dashed lines).}
\label{fig:40ms_chi_f}
\end{figure}

\subsection{Disruptive collision at 40 m/s}\label{sec:40ms}
Figure \ref{fig:collage}(c) depicts the outcome of a collision at $v_\mathrm{rel} = 40~\mathrm{m/s}$. At these higher impact speeds, the collision does not result in a perfect merger. Instead, a distribution of fragments is formed that in this case is characterized by three larger remnants and a tail of smaller ejecta made up of small aggregates and free-floating monomers. This outcome is consistent with -- and described in more detail in -- the study of \citet{Hasegawa2023}.

\begin{figure*}
\includegraphics[width=1.\textwidth]{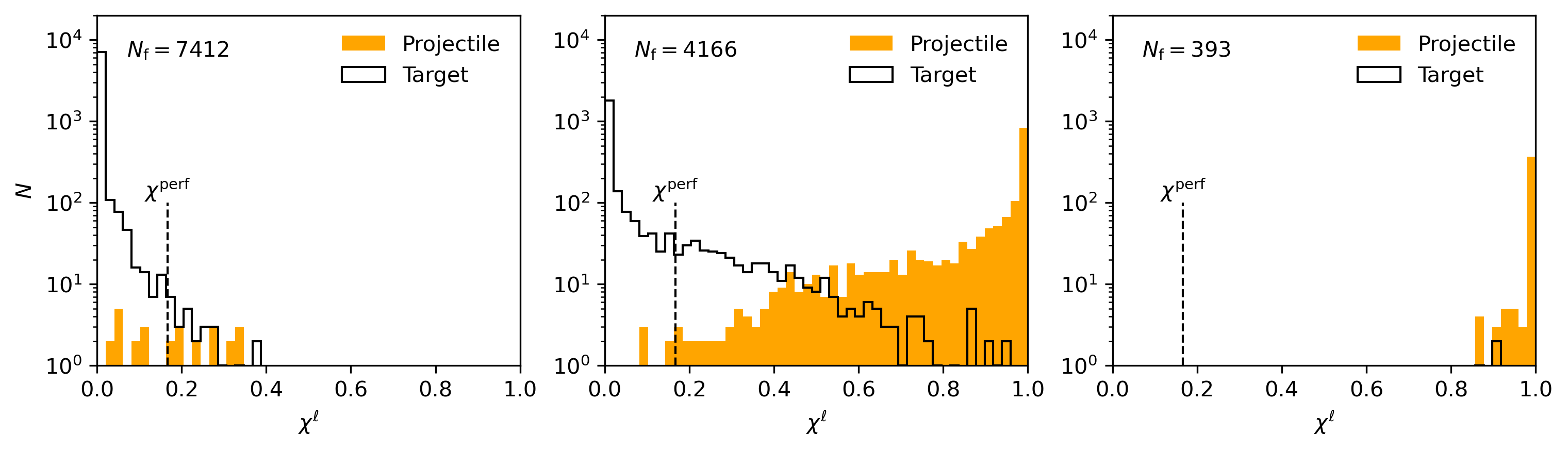}
\caption{Local projectile fraction for the largest (left), second-largest (middle), and third-largest (right) fragments shown in Fig.~\ref{fig:collage}(c), illustrating the disparity in compositions.\label{fig:40ms_chi_ell}}
\end{figure*}

\begin{figure*}
\includegraphics[width=1.\textwidth]{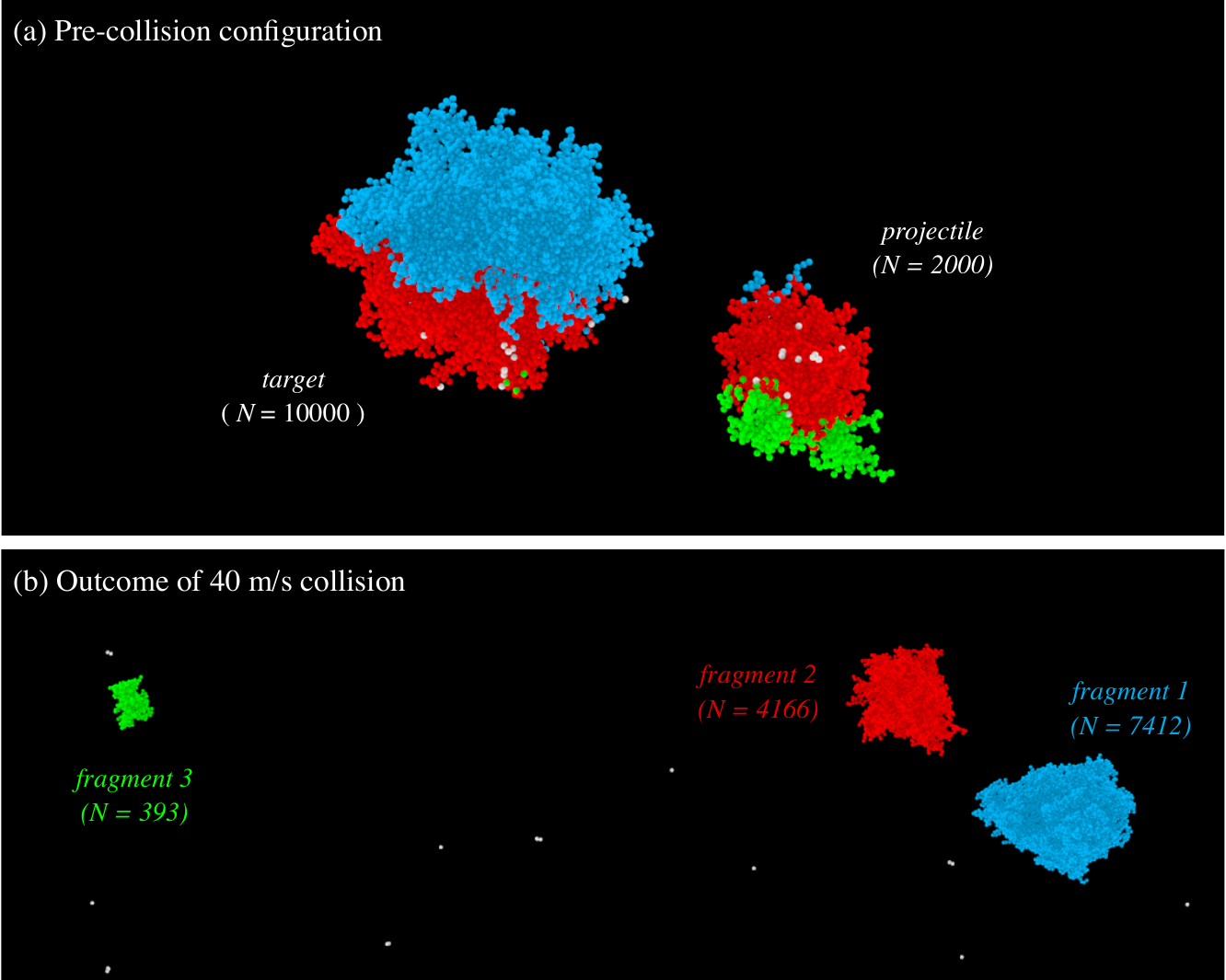}
\caption{Same as Fig.~\ref{fig:collage} panels (a) and (c) but with monomers color-coded according to their final position: blue monomers end up in the largest fragment, red monomers in the 2nd-largest fragment, green monomers in the 3rd-largest fragment, and white monomers are either free-floating or part of very small clusters after the 40 m/s impact.
\label{fig:collage2}}
\end{figure*}

To quantify the diversity in the fragments produced we show in Fig.~\ref{fig:40ms_chi_f} the masses, characteristic radii, and (macroscopic) projectile fractions for the 25 different fragments produced. Focusing on the three largest fragments\footnote{The smallest fragments -- individual monomers and clusters with $M_\mathrm{f} < 10m_\bullet$ -- together contain only $0.2\%$ of the total mass and we do not discuss them further here.} in the right half of the plot, what stands out is the compositional heterogeneity: the largest remnant consists of $0.5\%$ projectile material, the second-largest of $37\%$ projectile material, and the third-largest remnant consists for $99.2\%$ of material originally contained in the projectile. Individually, none of these come close to the $\chi^\mathrm{perf} = 1/6$ ratio corresponding to perfect mixing. The characteristic radii of the most massive fragments fall between those of the the original target and projectile. Internal projectile fractions for the three largest fragments are shown in Fig.~\ref{fig:40ms_chi_ell}. Finally, another way of illustrating the (lack of) mixing is presented in Fig.~\ref{fig:collage2}, which shows the same aggregates as Fig.~\ref{fig:collage} panels (a) and (c), but this time the monomers have been colour-coded according to which fragment they belong to \emph{after} the collision - making it possible to see which monomers end up where already in the pre-collision configuration.

\section{Collisions of BCCA Clusters}\label{sec:BCCA}
\begin{figure*}
\centering
\includegraphics[width=1.\textwidth]{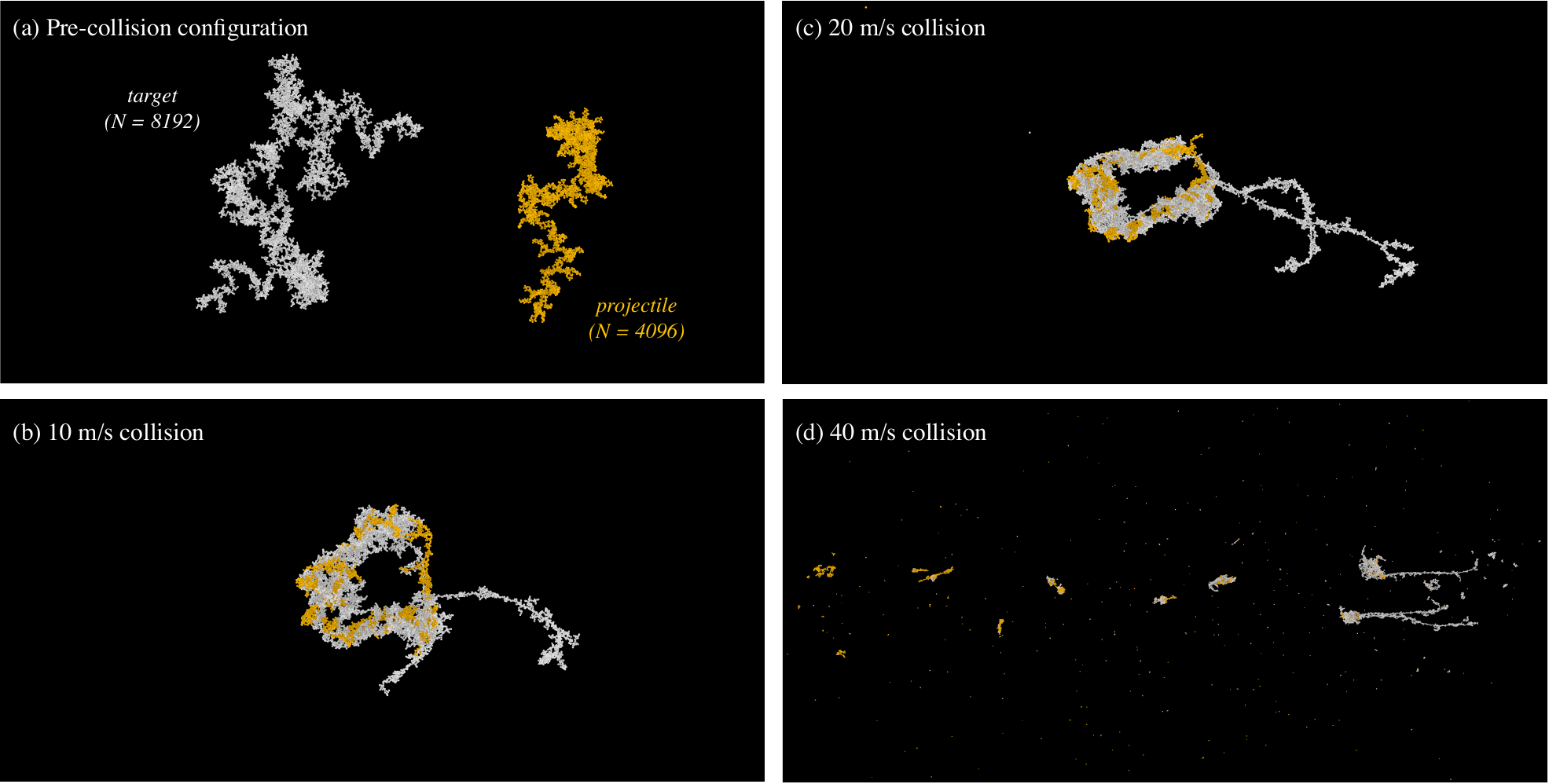}
\caption{(a) Pre-collision configuration showing the BCCA target (white) and projectile (gold) aggregates. (b)-(d) Outcomes of collisions $v_\mathrm{rel} = 10, 20, 40~\mathrm{m/s}$.
\label{fig:collage_bcca}}
\end{figure*}

In this Section we focus on collisions between BCCA clusters, which resemble more closely the highly-porous aggregates that are formed during the early `hit-and-stick' phase of dust coagulation in protoplanetary environments \citep[e.g.,][]{DominikTielens1997,blumwurm2008,okuzumi2009,Okuzumi2012,Krijt2015}. The target and projectile -- depicted in Fig.~\ref{fig:collage_bcca}(a) -- contain 8192 and 4096 monomers respectively, corresponding to a mass ratio of 1:2 and a perfect mixing ratio of $\chi^\mathrm{perf}=(1/3)$. We consider collision velocities of 10, 20, and 40 m/s, the outcomes of which are shown in Fig.~\ref{fig:collage_bcca}(b)-(d).

\begin{figure*}
\includegraphics[width=1.0\columnwidth]
{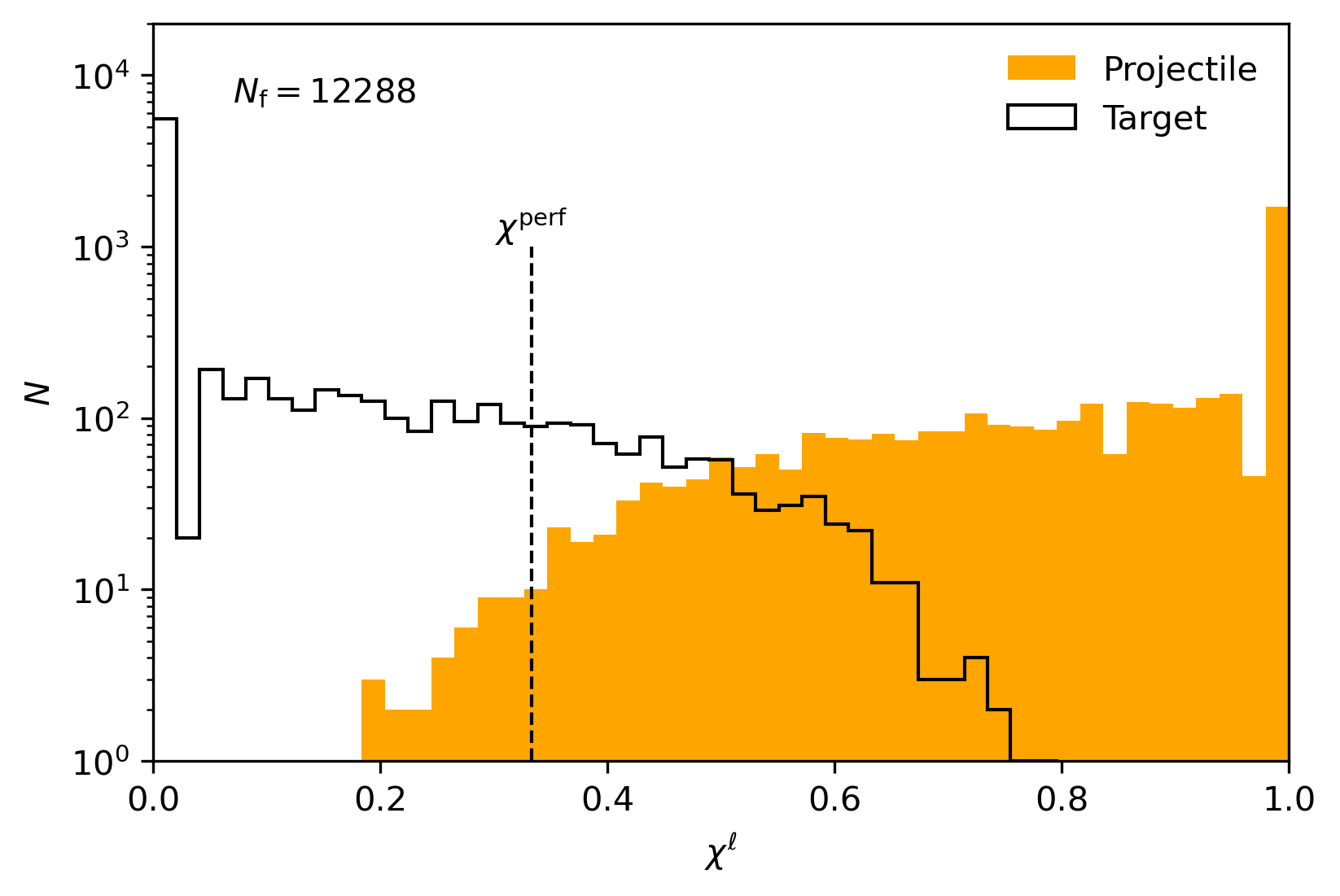}
\includegraphics[width=1.0\columnwidth]
{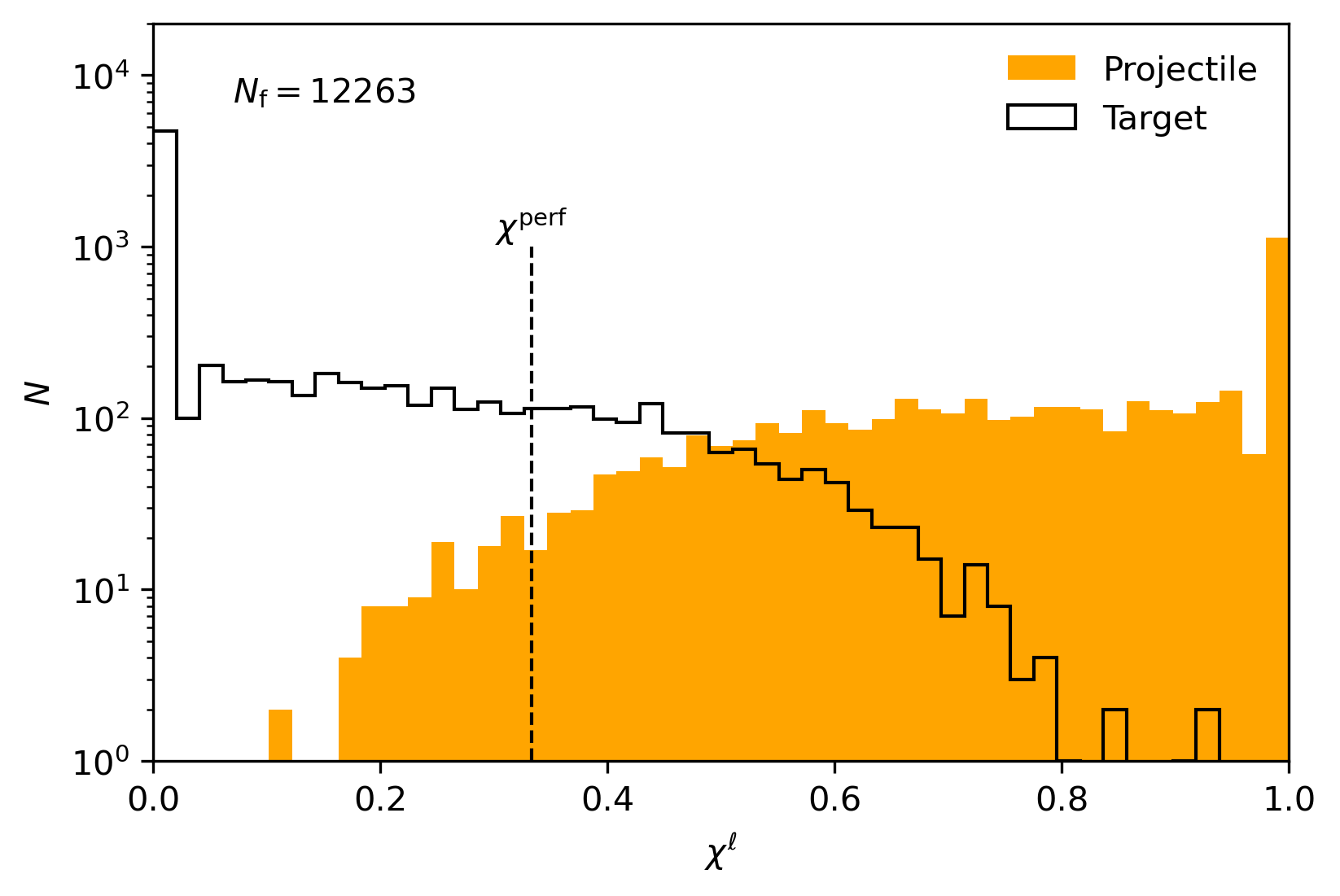}
\caption{Local projectile fraction for the collision products of BCCA aggregates colliding at 10 m/s (left) and 20 m/s (right).
\label{fig:10_20ms_bcca_chi_ell}}
\end{figure*}

\begin{figure*}
\includegraphics[width=1.0\columnwidth]
{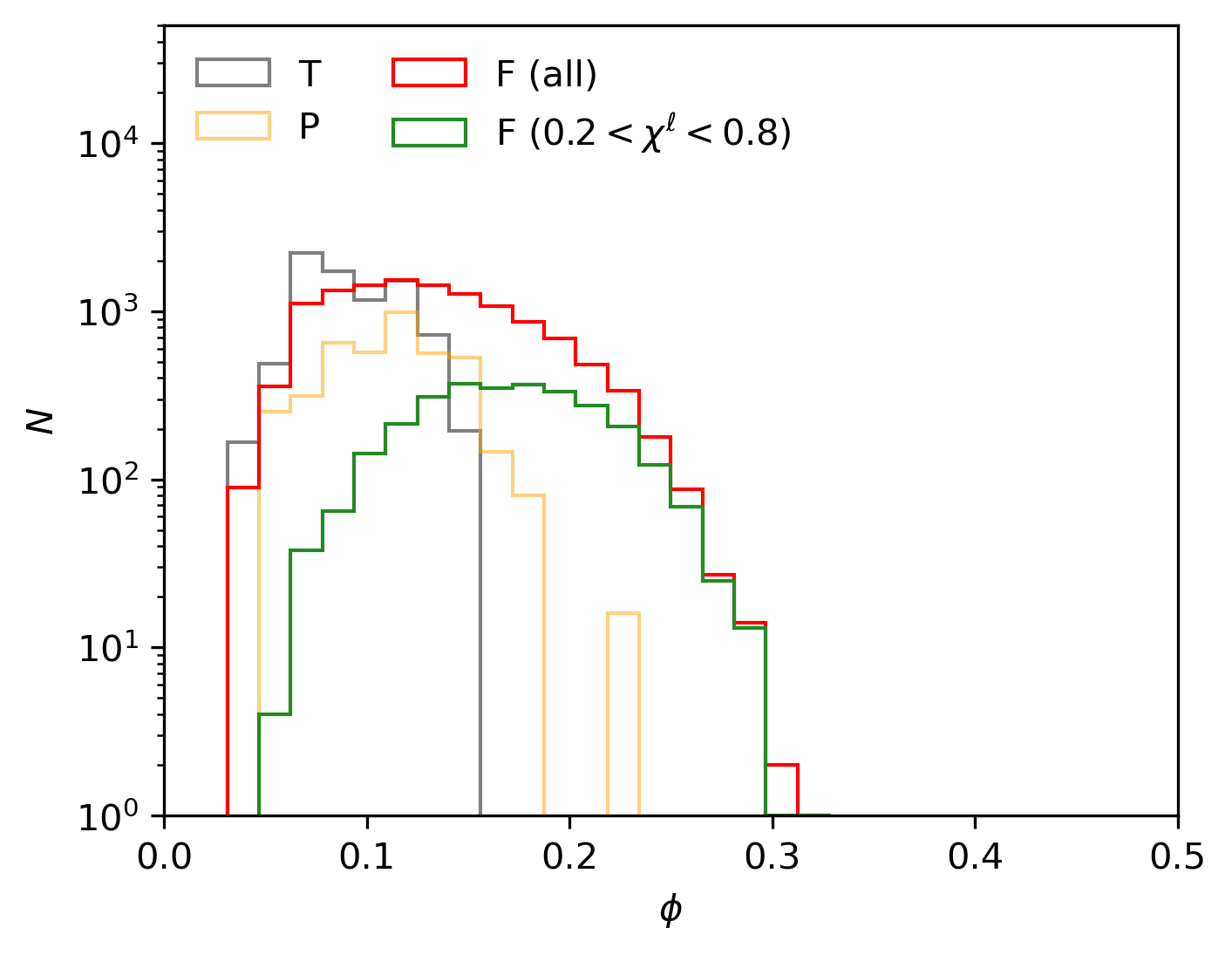}
\includegraphics[width=1.0\columnwidth]
{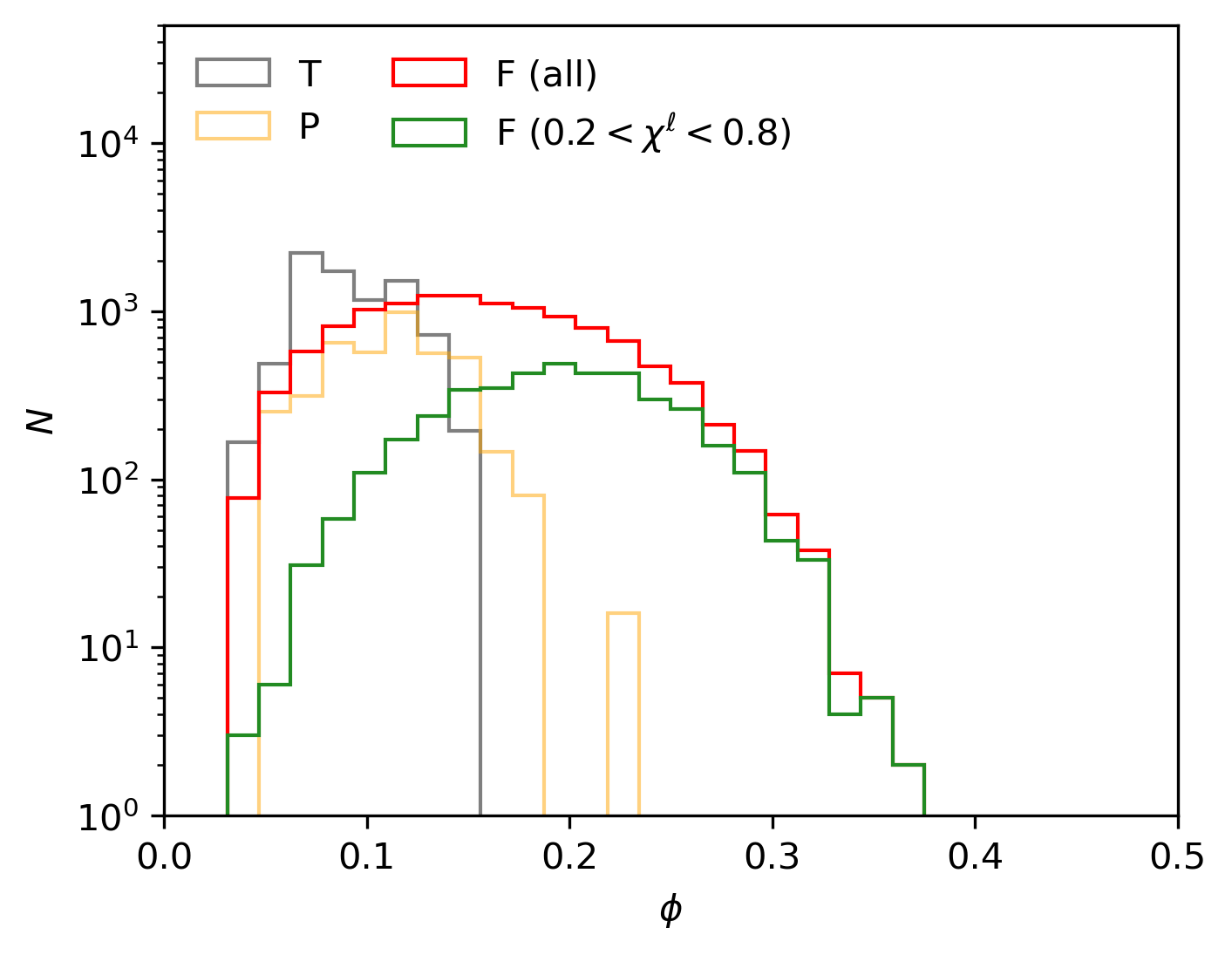}
\caption{Mass-weighted distributions of internal filling factor $\phi$ in the pre-collision BCCA target (T), BCCA projectile (P), the entire collision product (F, all), and the well-mixed portion of the fragment (F, $0.2 < \chi^\ell < 0.8$) for collisions at 10 m/s (left) and 20 m/s (right).
\label{fig:10_20ms_bcca_ff}}
\end{figure*}

\subsection{Sticking and restructuring at 10 and 20 m/s}

Both the 10 and 20 m/s collisions result in (near-) perfect sticking, with the final product containing 100\% and 99.8\% of the total mass, respectively. Inspection of Figs.~\ref{fig:collage_bcca}(b) and (c), especially when compared to the BPCA collision shown in Fig.~\ref{fig:collage}(b), suggests target and projectile materials are more readily mixed in the BCCA aggregate collision. This conclusion is supported by the local projectile fraction distributions, shown in Fig.~\ref{fig:10_20ms_bcca_chi_ell}, which show a considerable amount of material at intermediate $\chi^\ell$. It appears that the porous, open structure of the BCCA aggregates is more susceptible to restructuring, with kinetic energy being converted to rolling energy and the aggregates effectively folding in on themselves as they become more compact and mixed (at least at these relatively high velocities). We note however that the impact parameter and mass ratio also differed between the BCCA and BPCA impacts (see Sect.~\ref{sec:collisions_considered} which may contribute to the differences in outcomes. The compaction process is readily visible in the filling factor distributions shown in Fig.~\ref{fig:10_20ms_bcca_ff}, which indicate that, while the average local filling factor in the original BCCA aggregates is ${\approx} 0.1$, the newly created aggregate has an average $\phi \approx 0.2$. Compaction increases with collision velocity, with some regions reaching $\phi>0.3$ for the 20 m/s collision.

\begin{figure}
\includegraphics[width=1.0\columnwidth]
{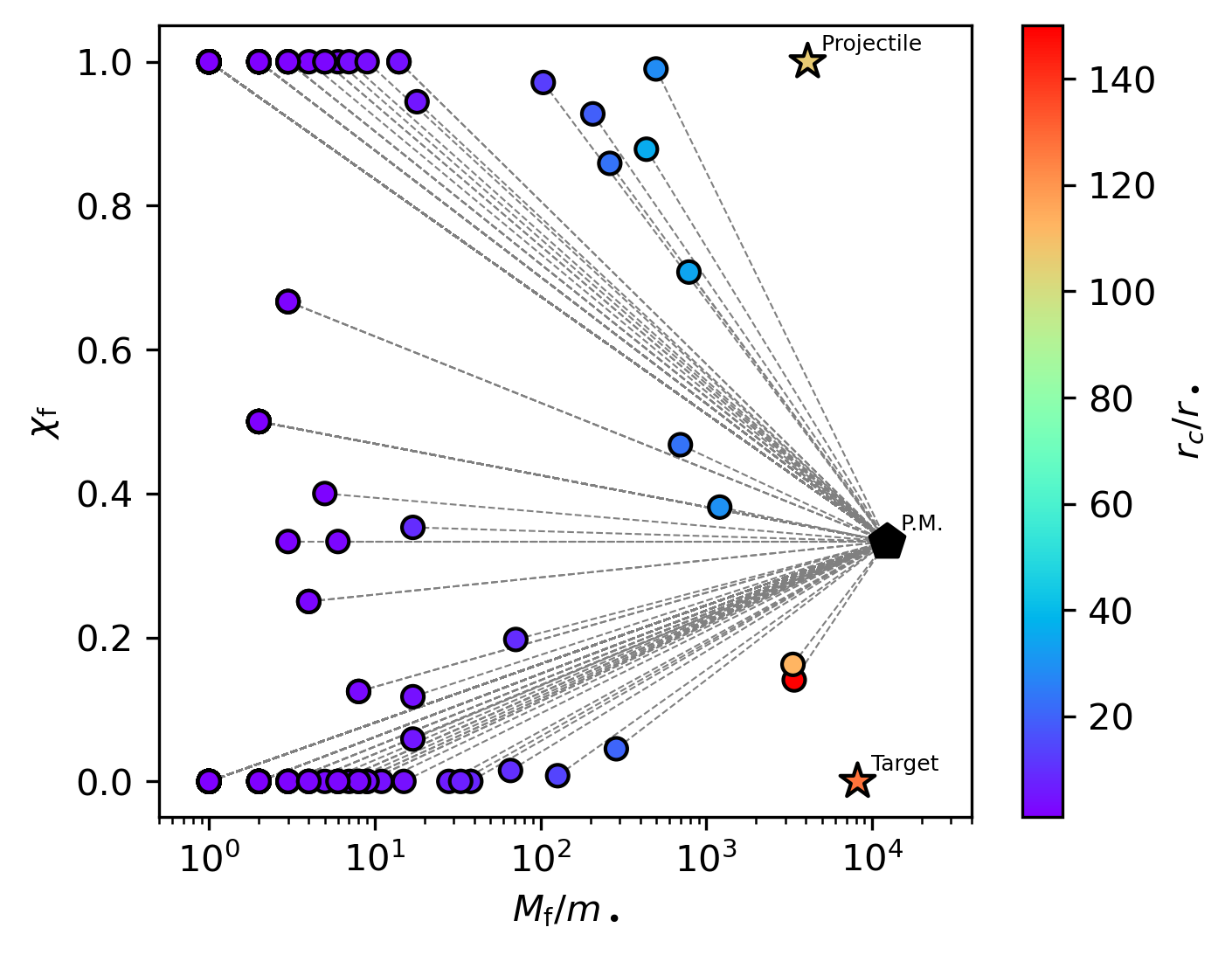}
\caption{Circles: Fragment masses, characteristic radii, and (macroscopic) projectile fractions ($\chi_\mathrm{f}$, see Sect.~\ref{sec:quantifying_mixing}) of the fragments produced in the 40 m/s BCCA collision depicted in Fig.~\ref{fig:collage_bcca}(d). 
\label{fig:40ms_bcca_chi_f}}
\end{figure}

\begin{figure*}
\includegraphics[width=1.\textwidth]{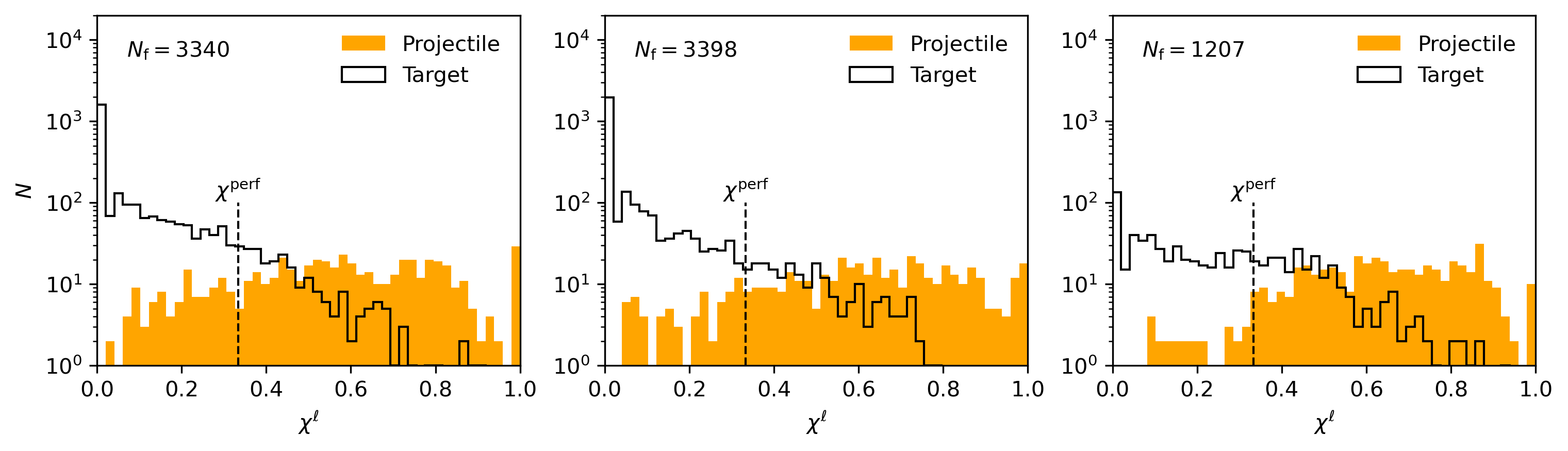}
\caption{Local projectile fraction for the largest (left), second-largest (middle) and third-largest (right) fragments created in the 40m/s BCCA collision shown in Fig.~\ref{fig:collage_bcca}(d).\label{fig:40ms_bcca_chi_ell}}
\end{figure*}

\subsection{Disruptive collision at 40 m/s}
At 40 m/s, even the porous BCCA aggregates are broken up as the increased kinetic energy results in the severing of large numbers of monomer-monomer bonds (Fig.~\ref{fig:collage_bcca}(d)). Compared to the BPCA collision at the same velocity (Fig.~\ref{fig:collage}(c)), more fragments are created and they appear to sample the full range of compositions. This is illustrated in Fig.~\ref{fig:40ms_bcca_chi_f}, where the most massive fragments (those consisting of ${>}100$) monomers show a wide diversity of macroscopic $\chi_\mathrm{f}$ values. The large characteristic radii of the more massive fragments are the result of their elongated shapes (see Fig.~\ref{fig:collage_bcca}). Figure~\ref{fig:40ms_bcca_chi_ell} shows the microscopic projectile fraction distributions for the three largest fragments (the only ones containing ${>}10^3$ monomers). Compared to the similar plot for compact aggregates (Fig.~\ref{fig:40ms_chi_ell}) we find that the BCCA collisions result in more internal mixing.

\section{Discussion}\label{sec:discussion}

We quantify and compare the effectiveness of internal mixing in the various collisions described in Table~\ref{tab:mixing_masses}. The columns here indicate the mass fraction that ends up in the `well-mixed' zone, defined either as having $(0.2 < \chi^\ell < 0.8)$ or $(0.5\chi^\mathrm{perf} < \chi^\ell < 2\chi^\mathrm{perf})$, with the latter option being more appropriate for collisions with high mass ratios. For the (offset) BPCA collisions, mass fractions in these zones range between ${\approx}3{-}6\%$, while the products of (head-on) BCCA collisions contain between ${\approx}19{-}33\%$ of material in the mixed region, an increase of a factor $5{-}10$.

These numbers suggest that single collisions are relatively inefficient ways of mixing, on small scales, materials of different origins (i.e., belonging to the target or projectile). The time it takes collisions to `mix' materials thoroughly can then be estimated as
\begin{equation}
    t_\mathrm{coll,mix} = \frac{t_\mathrm{coll}}{f_m} \simeq \frac{1}{f_m \Omega_K} \left( \frac{\Sigma_\mathrm{d}}{\Sigma_\mathrm{g}}\right)^{-1},
\end{equation}
where $f_m$ is the mass fraction mixed in a single collision (listed in Table~\ref{tab:mixing_masses}) and the collision timescale $t_\mathrm{coll}$ is written in terms of the dust-to-gas (surface density) ratio $\Sigma_\mathrm{d}/\Sigma_\mathrm{g}$ and the orbital frequency $\Omega_K$ following \citet[][Eq.~32]{birnstiel2023}. For typical values of $\Sigma_\mathrm{d}/\Sigma_\mathrm{g}=0.01$, $f_m = 0.05$, in a disc at a distance $r$ from a Sun-like star this results in $t_\mathrm{coll,mix} \approx 10,000~\mathrm{yr} ~(r/\mathrm{au})^{3/2}$, a considerable amount of time compared to other physical and chemical timescales in discs \citep[][Fig.~3]{semenov2011}. In a more complete description mixing may initially proceed quickly, as aggregates begin porous and $f_m \sim 0.2-0.3$, but then slow down substantially as successive collisions lead to compaction \citep[e.g.,][]{michoulier2024} and a reduction in $f_m$.

Another way of mixing monomers efficiently would be fully destroy aggregates and rebuild them from (diverse) monomers from the bottom up. However, the mass contained in the smallest sizes in typical coagulation/fragmentation steady state distributions is small \citep[e.g.,][]{birnstiel2011}, and furthermore coagulation is expected to be limited by radial drift -- not fragmentation -- in large parts of the outer disc regions \citep{birnstiel2023}.

If indeed largely heterogeneous aggregates persist for a considerable amount of time, this can lead to a variety of interesting effects: The sticking/fragmentation behaviour of aggregates, shaped by the composition of material near the aggregate surface \citep{musiolik2016}, may vary locally from aggregate to aggregate, or even for a single aggregate depending on from what side it is impacted. In the colder disc regions, grain surface chemistry and ice sublimation rates will be altered, as different parts of the aggregate's surface can be covered in different ice mantles. Finally, optical properties and the spectral appearance of aggregates will be affected \citep{min2008}, complicating the interpretation of spectroscopic observations. 

Future studies on a population/disc-wide level will have to assess in more detail how the mixing of different populations proceeds. This process is difficult to study with current grid-based methods, which typically do not include multiple grain compositions \citep[like DustPy, see][]{stammler2022} or have the built-in assumption that the composition -- at least locally -- is always constant across grain sizes \citep[e.g.,][]{boothilee2019, schneider2021}, while Monte Carlo methods \citep{Krijt2016, okamoto2022, HougeKrijt2023, Oosterloo2023, Oosterloo2024} or even SPH simulations \citep{michoulier2024} may provide a path forward.

On the individual collision level, an interesting follow-up to the work presented here, apart from a larger parameter study exploring larger variations in impact velocity, mass ratios, impact parameters, etc., would be to investigate multiple successive collisions; i.e., take the outcomes depicted in Fig.~\ref{fig:collage}(b-c) and use these as either target or projectile in the next collision. Along those lines, it would also be interesting to consider targets and projectiles made up of significantly different materials (e.g., with different surface energies, Young's modulus etc.) to see how the created interfaces between these different materials behave in successive impacts. If during fragmentation aggregates preferentially split apart again at these faults, the mixing of materials could be delayed further. Such studies may be challenging with current numerical set-ups, however, which typically assume a single monomer species \citep[e.g.,][and others]{Wada2013, Hasegawa2023}.

\begin{table}
    \centering
    
    \begin{tabular}{ l l | c c }
    \hline \hline
    Collision & $v_\mathrm{col}$ &  $(0.2 < \chi^\ell < 0.8)$  & $(0.5\chi^\mathrm{perf} < \chi^\ell < 2\chi^\mathrm{perf})$  \\
    \hline
       BPCA  & 20 m/s & 0.050  & 0.028 \\
        & 40 m/s$^*$  & 0.056 & 0.038 \\
        
       BCCA & 10 m/s  & 0.236 & 0.201 \\
       & 20 m/s  & 0.325 & 0.279 \\
       & 40 m/s$^*$  & 0.289  & 0.193 \\
\hline
       
    \end{tabular}
    \caption{Mass fraction (i.e., fraction of monomers) that ends up in a `well-mixed' zone, defined either as having $(0.2 < \chi^\ell < 0.8)$ (first column) or $(0.5\chi^\mathrm{perf} < \chi^\ell < 2\chi^\mathrm{perf})$ (second column). $^*=$ calculated using three largest fragments, which together contained 99.8\% (BPCA) and 65\% (BCCA) of the total mass involved in the collisions.}
    \label{tab:mixing_masses}
\end{table}

\section{Conclusions}

We have performed 3D simulations of individual aggregate-aggregate collisions (Figs. \ref{fig:collage}, \ref{fig:collage_bcca}) to study mixing between target and projectile material at both fragment and sub-aggregate levels. Our main conclusions can be summarized as follows:

\begin{enumerate}
    
    \item In sticking collisions of relatively compact (BPCA) aggregates, target and projectile do not readily mix, resulting in a sharp boundary in the resulting product (Figs. \ref{fig:20ms_chi_ell}, \ref{fig:20ms_interface}).
    
    \item The degree of mixing depends on the structure of the precursor bodies and collision geometry, however, with head-on collisions between BCCA aggregates resulting in increased mixing (Fig. \ref{fig:10_20ms_bcca_chi_ell}).

    \item In disruptive collisions at higher velocities, the larger fragments show a large compositional diversity, with few of them coming close to a projectile:target mass ratio that corresponds to a perfect merger (Figs. \ref{fig:40ms_chi_f}, \ref{fig:40ms_bcca_chi_f}).

    \item After individual collisions at similar impact velocities, the mass fraction of material that can be described as `well-mixed' increases from 3-6\% (for BPCA aggregates) to 20-30\% (for BCCA aggregates) (see
    Table~\ref{tab:mixing_masses}).
    
\end{enumerate}
Our findings imply that the homogenization of dust populations with different origins (which may appear naturally as the result of dynamical processes or thermal processing, see e.g., \citealt{Oosterloo2023, Oosterloo2024}) can take a substantial amount of time. During this time heterogeneous aggregates may be common and we have discussed some potential consequences of this in Sect.~\ref{sec:discussion}.

\section*{Acknowledgements}
We thank the anonymous referee for a constructive report and SA acknowledges support from JSPS KAKENHI Grant Number JP24K17118.


\section*{Data Availability}
Simulation output files used to generate the figures in this manuscript (e.g., text files containing each monomer's position at the beginning and end of the simulations) will be shared upon request.
 



\bibliographystyle{mnras}
\bibliography{references} 




\appendix

\section{Impact of the choice of $\ell$}\label{sec:appendix_ell}

Our definitions of the local projectile fraction $\chi^\ell$ (Sect.~\ref{sec:quantifying_mixing}) and filling factor $\phi$ (Eq.~\ref{eq:filling_factor}) involve defining a `local' volume with a radius $\ell$, the definition of which requires $r_\bullet < \ell < r_c$. In the main text we set $\ell = 5 r_\bullet$ which we found ensured decent numbers of local monomers while still allowing us to investigate sub-aggregate-level mixing. Here we briefly investigate the impact of this choice on our results by re-analyzing the outcome of the 20 m/s BPCA collision (Sect.~\ref{sec:20ms}) using $\ell = 3 r_\bullet$ and $\ell = 10 r_\bullet$. Average values for $N^\ell$ were 7, 26, and, 170, respectively, for increasing values of $\ell$.

\begin{figure}
\includegraphics[width=1.\columnwidth]{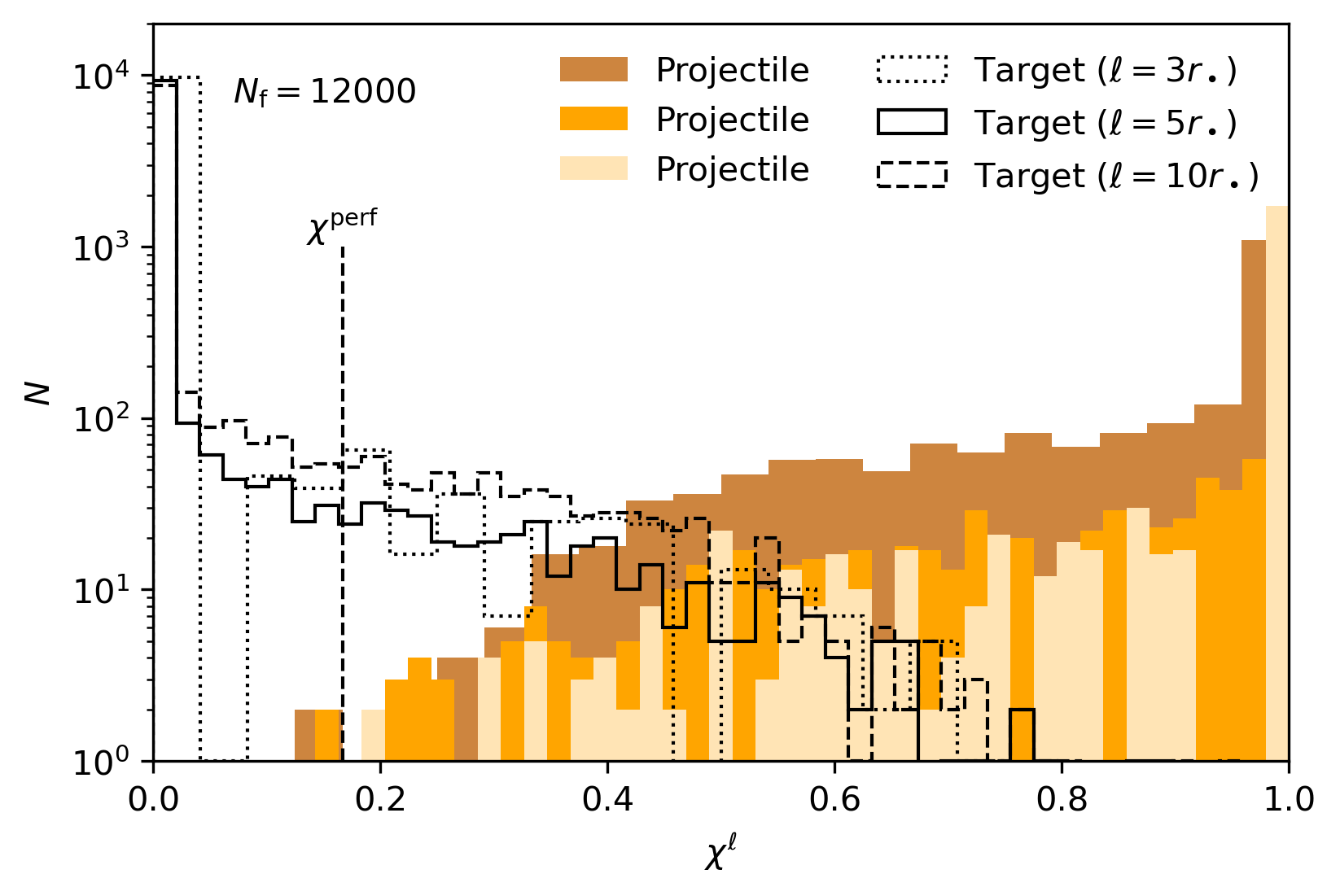}
\caption{Same as Fig.~\ref{fig:20ms_chi_ell} but with additional curves for $\ell=3r_\bullet$ and $10 r_\bullet$. Note that the bin width has been doubled for the $\ell=3r_\bullet$ distributions to reduce the scatter due to small number statistics.
\label{fig:appendix_chi}}
\end{figure}

\begin{table}
    \centering
    \begin{tabular}{ l c c | c c }
    \hline \hline
    Collision & $\ell$ & $\langle N^\ell \rangle$  &  $(0.2 < \chi^\ell < 0.8)$  & $(0.5\chi^\mathrm{perf} < \chi^\ell < 2\chi^\mathrm{perf})$  \\
    \hline
 BPCA   & $3 r_\bullet$ &  7 &0.029  & 0.018 \\
 $(20~\mathrm{m/s})$ & $5 r_\bullet$ & 26 & 0.050  & 0.028 \\
              & $10 r_\bullet$ &  170 & 0.093  & 0.053 \\
        
\hline
       
    \end{tabular}
    \caption{Impact of the choice of $\ell$ on the mass fraction that ends up the `well-mixed' zone (final two columns) for the 20 m/s BPCA collision introduced in Sect.~\ref{sec:20ms}. $\langle N^\ell \rangle$ represents the average number of monomers included when computing `local' filling factors and projectile fraction distributions.}   \label{tab:appendix}
\end{table}

Fig.~\ref{fig:appendix_chi} shows the local projectile fraction distributions for all three values of $\ell$. While the shapes are similar, the amount of mass with intermediate $\chi^\ell$ increases with increasing $\ell$, as is to be expected as increasingly distant neighbours are included in calculating the local composition. Table~\ref{tab:appendix} shows the obtained well-mixed mass fractions (see Sect.~\ref{sec:discussion}) for the 20 m/s BPCA collision as a function of $\ell$. The results suggest that the well-mixed fraction $f_m$ scales approximately linearly with the choice of $\ell$. This can be explained by the flat, plane-like geometry of the interface between the target-dominated and projectile-dominated parts of the formed aggregate (see Fig.~\ref{fig:20ms_interface}), resulting in the volume associated with this region growing linearly with $\ell$ as its thickness increases.

The local filling factor distribution (Fig.~\ref{fig:20ms_ff}) was also found to be somewhat sensitive to the choice of $\ell$. While the shapes remained similar, using $\ell=3r_\bullet$ resulted in the appearance of a small number (${\sim}10^2$) of counts with high filling factors $0.45 < \phi < 0.5$, while the larger $\ell=10r_\bullet$ instead shifted the filling factor distribution to lower values of $\phi$ and removed all regions with $\phi > 0.35$ as the larger `local' volumes began to smear out small-scale variations and included larger regions of empty space for monomers within $10 r_\bullet$ of the edge of the aggregate.

We conclude that while the observed behaviour is qualitatively the same, the exact values for the local filling factor and mixing ratios depend on the specific length-scales that are of interest.



\bsp	
\label{lastpage}
\end{document}